\documentclass[12pt]{article}
\usepackage{latexsym}
\usepackage{amsmath,amssymb}
\usepackage{bm}
\usepackage[dvipdfmx]{graphicx}
\usepackage[dvipdfmx]{xcolor}
\usepackage{amssymb}
\usepackage{cite}
\usepackage{amsfonts, amscd, amsthm}
\usepackage[all]{xy}
\usepackage{here}
\usepackage{ulem}
\usepackage{mathabx}

\renewcommand{\theequation}{\thesection.\arabic{equation}}

\makeatletter
\@addtoreset{equation}{section}
\renewcommand{\theequation}{\thesection.\@arabic\c@equation}
\makeatother

\makeatletter
\renewcommand\appendix{\par
  \setcounter{section}{0}%
  \setcounter{subsection}{0}%
  \gdef\thesection{Appendix \@Alph\c@section }
  \renewcommand{\theequation}
  {\Alph{section}.\arabic{equation}}
}
\makeatother

\makeatletter
\newcounter{subeqncnt}
\def\thesubeqncnt{\alph{subeqncnt}}
\def\subequations{\begingroup%
\stepcounter{equation}\edef\@tempa{\theequation}%
\let\c@equation\c@subeqncnt\c@subeqncnt\z@
\edef\theequation{\@tempa\noexpand\thesubeqncnt}}

\makeatother
\allowdisplaybreaks

\begin{document}

\titlepage

%%%%%%%%%%%%%%%%%%%%%%%%%%%%%%%%%%%%%%%%%%%%%%%%%%%%%%%%%%%%%%%%%%%%
\title{Two-Dimensional Kowalevski Top Problem\\ 
as the Genus Two Full Jacobi Inversion Problem\\
-Formulation Using Kolossoff Variables-} 
%%%%%%%%%%%%%%%
\author{
Kazuyasu Shigemoto\thanks{shigemot@tezukayama-u.ac.jp} \\
Tezukayama University, Nara 631-8501, Japan\\
}
\date{\empty}

%%%%%%%%%%%%%%%%%%%%%%%%%%%%%%%%%%%%%%%%%%%%%%%%%%%%%%%%%

\maketitle

\abstract{ 
For the Kowalevski top, we introduce another variable in addition to 
time variable, and formulate as the two-dimensional model.
In this paper, since the results for the genus two full Jacobi inversion problem for the two 
variables Kowalevski top are already known, we utilize these results in reverse, 
introducing Kolossoff variables, to derive a more transparent formulation of the 
genus two full Jacobi inversion problem.
Since there are two implicit equations of motion using Kowalevski variables, 
we derived two explicit equations of motion using Kolossoff variables,
which is different from the Kolossoff method to derive only one explicit 
equation and to use Hamiltonian-Jacobi theory.
}
\section{Introduction} 
\setcounter{equation}{0}

Weierstrass $\wp$-function, which is the solution of the static KdV equation, is 
generalized into higher genus hyperelliptic $\wp$-functions.
Since these can be described using a single
 potential function $\log \sigma$ by using the 
$\sigma$-function~\cite{Baker1, Baker2}, they give
 rise to a family of higher-dimensional integrable systems.
A typical example of a dynamical model involving hyperelliptic $\wp$-functions
with genus two is the Kowalevski top \cite{Kowalevski,Kotter,Kolossoff}.
It has been known for over a century that the Kowalevski top is exactly solvable; however, 
it was later demonstrated that it constitutes an integrable system via a Lax 
pair \cite{Perelomov1, Reyman1}. Furthermore, it was shown that the 
Kowalevski top possesses
 the structure of the Lie group $SO(3,2)$ \cite{Reyman1}. Additionally, equations of motion
 governing the time evolution with respect to another variable have also been
 discovered, that is, Kowalevski top is generalized to the 
two-dimensional model~\cite{Haine, Adler}. 
Thus, new developments continue to emerge to this day
~\cite{Reyman2,Marshall,Markushevich,Perelomov2,Sokolov,Dragovic,Hayashi5}.

In the original Kowalevski top, there is only a single time 
variable $t$. Consequently, although the solutions involve genus two 
hyperelliptic $\wp$-functions, 
they depend on only one argument rather than two arguments typically associated with 
such hyperelliptic functions. To address this, in our previous 
work \cite{Hayashi5}, we introduced another
 variable $r$, and formulated as the two-dimensional Kowalevski top. 
By considering evolution of the Kowalevski top with respect to 
$t$ and $r$, we were able to derive a formulation of the genus two full Jacobi inversion problem.  

In this paper, we introduce Kolossoff variables \cite{Kolossoff} as an intermediate stage between
 the Kowalevski variables and the variables of the genus two full Jacobi inversion problem.
We demonstrate that
 this approach renders the derivation process clearer and more straightforward.

In a previous paper~\cite{Hayashi5}, by employing a Lax-pair formulation, it was shown that 
two-dimensional Kowalevski top is exactly solvable because the model possesses 
the structure of the 
Lie group $SO(3,2) \cong Sp(4, \mathbb{R})/\mathbb{Z}_2$. Since the two-dimensional  
Kowalevski top problem can be identified with the genus two full Jacobi inversion 
problem, the task of investigating the symmetries of the two-dimensional Kowalevski top is reduced 
to the more tractable problem of examining the structure of the genus two hyperelliptic 
$\wp$-functions. In earlier works~\cite{Hayashi6, Hayashi7,Hayashi8}, it was demonstrated that 
the structure of this genus two hyperelliptic $\wp$-functions is 
$SO(3,2) \cong Sp(4, \mathbb{R})/\mathbb{Z}_2$.

%%%%%%%%%%%%%%%%%%%%%%%%%%%%%%%%%%%%%%%%%%%%%%%%%%%%%%%%%%%%%%%%%%%%%%%%%%
\section{Time Evolving Kowalevski Top}
\setcounter{equation}{0}

Using the notation from a previous paper~\cite{Hayashi5}, we denote standard 
time derivatives as follows: $\partial_t \omega_1 = \dot{\omega}_1$, etc.
In this notation, the original equations of motion for the Kowalevski top~\cite{Kowalevski, Kotter}
 are given as follows:

\begin{align}
&2 \dot{\omega}_1=\omega_2 \omega_3 ,  \quad
2 \dot{\omega}_2=-\omega_3 \omega_1-c_0\gamma_3  ,  \quad
\dot{\omega}_3=c_0 \gamma_2  ,
\label{2e1}\\
&\dot{\gamma}_1=\omega_3 \gamma_2-\omega_2 \gamma_3, \quad
\dot{\gamma}_2=\omega_1 \gamma_3-\omega_3 \gamma_1 , \quad
\dot{\gamma}_3=\omega_2 \gamma_1-\omega_1 \gamma_2 . 
\label{2e2}
\end{align}
Here, $c_0$ is a constant.

Now, we have four conserved quantities $\{ E, \ell, 1, k\}$ as follows:
\begin{align}
&2E=2\omega_1^2+2\omega_2^2+\omega_3^2-2 c_0\gamma_1, \quad
2\ell=2\omega_1 \gamma_1+2\omega_2 \gamma_2+\omega_3 \gamma_3, \quad
1=\gamma_1^2+\gamma_2^2+\gamma_3^2, 
\label{2e3}\\
&k^2=\left[c_0(\gamma_1+i \gamma_2) + \left(\omega_1+i\omega_2\right)^2\right]
\left[c_0(\gamma_1-i \gamma_2) + \left(\omega_1-i\omega_2\right)^2\right] .
\label{2e4}
\end{align}
In other words, there are six variables $\omega_1$, $\omega_2$, $\omega_3$, 
$\gamma_1$, $\gamma_2$, $\gamma_3$  and there are four conserved
 quantities, which results in two independent equations of motion expressed in implicit form.

%%%%%%%%%%%%%%%%%%%%%%%%%%%%%%
\subsection{Kowalevski Variable}
We rewrite the equations of motion by introducing the following Kowalevski 
variables $x_1$ and $x_2$
\begin{align}
x_1=\omega_1+i \omega_2, \quad x_2=\omega_1-i\omega_2 .
\label{2e5}
\end{align}
Furthermore, we introduce variables $\xi_1, \xi_2, \eta_1, \eta_2$, which are 
not independent of $x_1$ and $x_2$ but they are nonetheless useful, as follows:
\begin{align}
&\xi_1=x_1^2+c_0 (\gamma_1+i \gamma_2), \quad 
\xi_2=x_2^2+c_0 (\gamma_1-i \gamma_2),
\label{2e6}\\
&\eta_1=\omega_3 x_1 +c_0 \gamma_3, \quad
\eta_2=\omega_3 x_2 +c_0 \gamma_3.  \quad 
\label{2e7}
\end{align}
Using these variables, Kowalevski's original equations of motion can be written as follows:
\begin{align}
&\dot{x_1}=-\frac{i}{2} \eta_1,  \quad
\dot{x_2}=\frac{i}{2} \eta_2  ,  \quad
\dot{\xi_1}=-i \xi_1 \omega_3 , \quad
\dot{\xi_2}=i \xi_2 \omega_3 , 
\label{2e8}\\
&\dot{\eta_1}=\frac{i}{2} \left\{ (\xi_1-x_1^2) (x_2-x_1)
-\omega_3 \eta_1\right\}, \quad
\dot{\eta_2}=-\frac{i}{2} \left\{(\xi_2-x_2^2) (x_1-x_2)
-\omega_3 \eta_2 \right\} .
\label{2e9}
\end{align}
The conserved quantity can also be expressed as follows: 
\begin{align}
&E=\left\{(x_1+ x_2)^2 +\omega_3^2-\xi_1-\xi_2\right\}/2, \quad
\gamma_3=\sqrt{1-\gamma_1^2-\gamma_2^2}, \quad
k=\sqrt{\xi_1 \xi_2}, \quad
\label{2e10}\\
&\ell=\left\{x_1 (\gamma_1-i\gamma_2)+x_2  (\gamma_1+i\gamma_2)+\omega_3 \gamma_3\right\}/2 .
\label{2e11}
\end{align}
In the case of $\gamma_3$, 
we give the expression in the form to solve for $\gamma_3$ by using the conservation law.

There are six variables $\{ x_1, x_2, \xi_1, \xi_2, \eta_1, \eta_2 \}$ and four conserved 
quantities $\{ E, \ell, 1, k \}$. Consequently, Kowalevski chose $x_1$ and $x_2$ as independent 
variables, and the equations for the Kowalevski top take the form of two implicit equations.
%%%%%%%%%%%%%%%%%%%%%%%%%%%%%%%%%%
\subsection{Kolossoff Variables}
Next, we define the Kolossoff variables as follows:
\begin{align}
x=\frac{(\xi_1 \eta_2^2+\xi_2 \eta_1^2)}{k(x_2-x_1)^2}, \quad 
y=-\frac{i(\xi_1 \eta_2^2-\xi_2 \eta_1^2)}{k(x_2-x_1)^2}  .
\label{2e12}
\end{align}
Consequently, the two implicit equations of motion in terms of Kowalevski variables 
$x_1$ and $x_2$ become two explicit equations of motion in terms of the Kolossoff 
variables $x$ and $y$. This is why the proof becomes complicated 
but straightforward when the 
Kolossoff variables are used.
The two independent equations of motion using Kolossoff variables are as follows:
\begin{align}
&\frac{1}{2}\left( \dot{x}^2+\dot{y}^2 \right)
-\left\{\rho^3-E \rho^2+(c_0^2-k x)\rho \right\}=0, 
\label{2e13}\\
&\frac{1}{2(x^2+y^2)}\left\{ (x^2-y^2-2 k x) \left(\dot{x}^2-\dot{y}^2\right)
 +4 (x-k) y  \dot{x} \dot{y} \right\}
\nonumber\\
&-\left\{ \rho^3 -E \rho^2+(c_0^2+2 k^2-3 k x)\rho+E (2 k x-2c_0^2)
+4 \ell^2 c_0^2 \right\}=0, 
\label{2e14}
\end{align}
where $\rho=\sqrt{x^2+y^2}=-2\eta_1 \eta_2/(x_2-x_1)^2(>0)$.
There are implict two equations of motion with Kowalevski variables.
While Kolossoff derived only one equation of motion Eq.(\ref{2e13}), but we derived two 
equations of motion Eq.(\ref{2e13}) and Eq.(\ref{2e14}) with Kolossoff variables.
Although this calculation is complicated, it is straightforward; its validity can 
be confirmed by using the 
the equations of motion Eq.(\ref{2e8}) and Eq.(\ref{2e9}) and the conservation 
laws Eq.(\ref{2e10}) 
and Eq.(\ref{2e11}). The expectation that the above equations 
should hold stems from the result of our previous paper~\cite{Hayashi5}, in which 
we derived the 
form of the genus two full Jacobi inversion problem. We then worked backward 
from that Jacobi inversion problem to deduce the above corresponding 
form Eq.(\ref{2e13}) and Eq.(\ref{2e14}). 
We subsequently verified that this expectation is indeed correct by explicitly 
applying Eqs.(\ref{2e8})-(\ref{2e11}).

The more detailed way to verify the above equations is given as follows: 
first, the derivatives of Kolossoff variables 
are expressed in terms of the derivatives of the Kowalevski variables 
using Eq.(\ref{2e8}) and Eq.(\ref{2e9}). Furthermore, by substituting 
the conserved quantities 
Eqs.(\ref{2e10}) and Eq.(\ref{2e11}), and employing the variables 
$\{\xi_1, \xi_2, \eta_1, \eta_2\}$ as the expression 
of $\{x_1, x_2, \gamma_1, \gamma_2, \omega_3\}$
using $\gamma_3=\sqrt{1-\gamma_1^2-\gamma_2^2}$, 
the above equations are expressed entirely in terms of the 
set $\{x_1, x_2, \gamma_1, \gamma_2, \omega_3\}$. 
By thus returning to the variables $\{x_1, x_2, \gamma_1, \gamma_2, \omega_3\}$, it has 
been confirmed that Eq.(\ref{2e13}) and Eq.(\ref{2e14}) hold true automatically.

%%%%%%%%%%%%%%%%%%%%%%%%%%%%%%%%
\subsection{Genus Two Full Jacobi Inversion Variables}
From Kolossoff variables $x$ and $y$, we define the variables $\lambda$ 
and $\mu$ of the genus two full Jacobi inversion problem as follows:
\begin{align}
x=(\lambda \mu  +k^2)/k, \quad y=\sqrt{(\lambda^2-k^2)(k^2-\mu^2)}/k . 
\label{2e15}
\end{align}
Since $x^2+y^2=(\lambda+\mu)^2$, if we choose the sign of the square root as 
$\rho=\sqrt{x^2+y^2}=-(\lambda+\mu)$, then the following relations hold
\begin{align} 
\lambda+\mu=-\sqrt{x^2+y^2}=-\rho, \quad \lambda \mu=k x-k^2 .
\label{2e16}  
\end{align}
In this case, the kinetic term in Eq.(\ref{2e13}), expressed in terms of Kolossoff variables, 
can be rewritten using the variables of the genus two full Jacobi inversion problem as follows:
\begin{align}
\dot{x}^2+\dot{y}^2=(\lambda^2-\mu^2)
\left( \frac{\dot{\lambda}^2}{\lambda^2-k^2}- \frac{\dot{\mu}^2}{\mu^2-k^2}\right) .
\label{2e17}
\end{align}
Furthermore, we introduce Kowalevski's $\varphi$ function as follows:
\begin{align}
\varphi(\lambda)=
\lambda^3+E \lambda^2+(c_0^2-k^2) \lambda+E(c_0^2-k^2)-2\ell^2 c_0^2 .
\label{2e18}
\end{align}
Using this expression, the following relationship can be shown
\begin{align}
\rho^3-E\rho^2+(c_0^2-k x) \rho
=-\frac{\lambda+\mu}{\lambda-\mu} 
\left(\varphi(\lambda)-\varphi(\mu) \right) .
\label{2e19}
\end{align}
Therefore, using Eqs.(\ref{2e13}), (\ref{2e17}) and (\ref{2e19}), we obtain the following 
equation 
\begin{align}
-\frac{1}{2} (\lambda-\mu)^2 
\left( \frac{\dot{\lambda}^2}{\lambda^2-k^2}- \frac{\dot{\mu}^2}{\mu^2-k^2}\right)
=\varphi(\lambda)-\varphi(\mu) .
\label{2e20}
\end{align}
Similarly, the kinetic term in Eq.(\ref{2e14}), expressed in terms of Kolossoff variables, 
can be rewritten using the variables of the genus two full Jacobi inversion problem as follows:
\begin{align}
(x^2-y^2-2 k x)(\dot{x}^2-\dot{y}^2)+4 (x-k) y \dot{x} \dot{y}
=(\lambda^2-\mu^2)^2
\left( \frac{\dot{\lambda}^2}{\lambda^2-k^2}+ \frac{\dot{\mu}^2}{\mu^2-k^2}\right) .
\label{2e21}
\end{align}
While, the following relationship can be shown
\begin{align}
\rho^3-E\rho^2+(c_0^2++2 k^2-3 k x) \rho+E(2 k x -2 c_0^2)+4 \ell^2 c_0^2 .
=-\left(\varphi(\lambda)+\varphi(\mu) \right) .
\label{2e22}
\end{align}
Therefore, using Eqs.(\ref{2e14}), (\ref{2e21}) and (\ref{2e22}), the following 
equation is obtained
\begin{align}
-\frac{1}{2} (\lambda-\mu)^2 
\left( \frac{\dot{\lambda}^2}{\lambda^2-k^2}+ \frac{\dot{\mu}^2}{\mu^2-k^2}\right)
=\varphi(\lambda)+\varphi(\mu) .
\label{2e23}
\end{align}
Thus, from Eq.(\ref{2e20}) and Eq.(\ref{2e23}), the following equations are obtained
\begin{align}
-\frac{1}{2} (\lambda-\mu)^2 \frac{\dot{\lambda}^2}{\lambda^2-k^2}
=\varphi(\lambda), \quad
-\frac{1}{2} (\lambda-\mu)^2 \frac{\dot{\mu}^2}{\mu^2-k^2}
=\varphi(\mu). 
\label{2e24}
\end{align}
By choosing the sign of the square root as shown below, we obtain the following equation
\begin{align}
\frac{\dot{\lambda}}{\sqrt{\varphi(\lambda)(\lambda^2-k^2)}}
=\frac{\sqrt{2} i}{\lambda-\mu}, \quad
\frac{\dot{\mu}}{\sqrt{\varphi(\mu)(\mu^2-k^2)}}
=\frac{\sqrt{2} i}{\mu-\lambda}=-\frac{\sqrt{2} i}{\lambda-\mu} .
\label{2e25}
\end{align}
This expression will be used later when discussing the genus two full Jacobi inversion problem. 

Kolossoff uses only Eq.(\ref{2e20}) and applies Hamilton-Jacobi theory, along 
with a change of variables, to derive the two relations presented in Eq.(\ref{2e24}).
However, since Eq.(\ref{2e20}) is not a separable differential equation, there is a logical gap 
in Kolossoff approach, which attempts to extract two distinct relations from a single equation. 
Consequently, Kolossoff original method gives the constant term 
for $\varphi$ in Eq.(\ref{2e18}) to be an arbitrary constant; it fails to specify the constant term in the 
more precise form $E(c_0^2-k^2)-2 \ell^2 c_0^2$.
In contrast, our approach avoids Hamilton-Jacobi theory; instead, we derive two independents 
equations of motion and solve them to arrive at Eq.(\ref{2e24}), 
which is the logically clearer method.

%%%%%%%%%%%%%%%%%%%%%%%%%%%%%%%%%%%%%%%%%%%%%%%%%%%%%%%%%%%%%%%%%%%%%%%%%
\section{Another Variable Evolving Kowalevski Top}
\setcounter{equation}{0}
Next, we consider the equations of motion along another variable $r$,
which are given as follows~\cite{Haine,Adler}:
\begin{align}
2\omega_1'&=-c_0 \gamma_2 \gamma_3 
+\omega_3 (\gamma_1 \omega_2-\gamma_2 \omega_1)
-2 \gamma_3 \omega_1 \omega_2
-\frac{1}{c_0} (\omega_1^2+\omega_2^2) \omega_2 \omega_3  , 
\label{3e1}\\
2\omega_2'&=c_0 \gamma_1 \gamma_3 
+\omega_3 (\gamma_1 \omega_1+\gamma_2 \omega_2)
+\gamma_3(\omega_1^2- \omega_2^2)
+\frac{1}{c_0} (\omega_1^2+\omega_2^2) \omega_1 \omega_3  , 
\label{3e2}\\
\omega_3'&=\gamma_2(\omega_1^2- \omega_2^2)
-2 \gamma_1 \omega_1 \omega_2 , 
\label{3e3}\\
\gamma_1'&=\gamma_3
\left(\gamma_1 \omega_2-\gamma_2 \omega_1
-\frac{1}{c_0} (\omega_1^2+\omega_2^2) \omega_2\right)  , 
\label{3e4}\\
\gamma_2'&=\gamma_3 
\left(\gamma_1 \omega_1+\gamma_2 \omega_2
+\frac{1}{c_0} (\omega_1^2+\omega_2^2) \omega_1\right)  , 
\label{3e5}\\
\gamma_3'&=-\omega_2(\gamma_1^2+\gamma_2^2)
+\frac{1}{c_0}(\gamma_1 \omega_2-\gamma_2 \omega_1)(\omega_1^2+\omega_2^2) .   
\label{3e6}
\end{align}
Under these equations of motion of another variable, the four 
conservation laws Eq.(\ref{2e3}) and Eq.(\ref{2e4}) are also conserved. Furthermore, 
both the ordinary time and another variable evolution are integrable; that is, 
the integrability conditions for $\omega_i$ and $\gamma_i$ ($i=1, 2, 3$) are satisfied.
For example, the integrability condition 
$\partial_ r \partial_t \omega_1= \partial_t \partial_ r \omega_1$ is equivalent to   
\begin{equation}
\partial_r (\omega_2 \omega_3)
=\partial_t \left(-c_0 \gamma_2 \gamma_3 
+\omega_3 (\gamma_1 \omega_2-\gamma_2 \omega_1)
-2 \gamma_3 \omega_1 \omega_2
-\frac{1}{c_0} (\omega_1^2+\omega_2^2) \omega_2 \omega_3 \right).
\label{3e7}
\end{equation}
It can be observed that Eq.(\ref{3e7}) holds by using the ordinary time evolving equations
Eq.(\ref{2e1}) and Eq.(\ref{2e2}) and another variable evolving equations 
Eqs.(\ref{3e1})-(\ref{3e6}).
Similarly, the integrability conditions for other variables are also satisfied.

%%%%%%%%%%%%%%%%%%%%%%%%%%%%%%%
\subsection{Kowalevski Variables}
First, we rewrite the equation of motion of another variable. Using Kowalevski variables 
$x_1$ and $x_2$, we obtain as follows:
\begin{align}
&x_1'=\frac{i}{2c_0} \xi_1 \eta_2, \quad x_2'=-\frac{i}{2c_0} \xi_2 \eta_1, \quad
\xi_1'=\frac{i}{2c_0} \xi_1 \left\{ (\eta_1+\eta_2)(x_1+x_2)-\omega_3(x_1^2+x_2^2) \right\} , 
\label{3e8}\\
&\xi_2'=-\frac{i}{2c_0} \xi_2 \left\{ (\eta_1+\eta_2)(x_1+x_2)-\omega_3(x_1^2+x_2^2) \right\}, 
\label{3e9}\\
&\eta_1=\frac{i}{2c_0} \xi_1 \left\{ (x_2-x_1)(x_2^2-\xi_2)+\omega_3 \eta_2 \right\}, 
\eta_2=-\frac{i}{2c_0} \xi_2 \left\{ (x_1-x_2)(x_1^2-\xi_1)+\omega_3 \eta_1 \right\} .
\label{3e10}
\end{align}
%

%%%%%%%%%%%%%%%%%%%%%%%%%%%%%%%%%
\subsection{Kolossoff Variables}
Using Kolossoff variables of Eq.(\ref{2e12}), the new two explicit independent 
equations of motion are given as follows:
\begin{align}
&\frac{c_0^2}{2 k^2 (x-k)^2}\left\{ \left((x-k)x'+y y' \right)^2 +k^2 y'^2 \right\}
-\left\{ \rho^3-E \rho^2+(c_0^2-k x) \rho \right\}=0 ,
\label{3e11}\\
&\frac{c_0^2}{2 k^2 (x^2+y^2) (x-k)^2}
\left[ (x-k)^2 (x^2+y^2-2 k x) x'^2 +2 (x-k)y(x^2+y^2-2 k x +2 k^2) x' y' \right.
\nonumber\\
&\left. +\left\{ (y^2+k^2) (x^2+y^2-2 k x+2 k^2) -2 k^2 (x-k)^2 \right\} y'^2 \right]
\nonumber\\
&-\left\{ \rho^3-E \rho^2+(c_0^2+2 k^2-3 k x) \rho 
+E(2 k x -2 c_0^2)+ 4 \ell^2 c_0^2 \right\}=0 .
\label{3e12}
\end{align}
Although this calculation is complicated, it is straightforward; its validity can 
be confirmed by using the 
Eqs.(\ref{3e8})-(\ref{3e10}), Eq.(\ref{2e10}) and Eq.(\ref{2e11}). The expectation 
that the above equations 
should hold stems from the result of our previous paper~\cite{Hayashi5}, in 
which we derived the 
form of the genus two full 
Jacobi inversion problem. We then worked backward 
from the genus two full Jacobi inversion problem to deduce the corresponding 
form of the above 
equations. We subsequently verified that this expectation is indeed correct by 
explicitly applying 
Eqs. (\ref{3e8})-(\ref{3e10}), Eq.(\ref{2e10}) and Eq.(\ref{2e11}).

The more detailed way to verify the above equations is given as follows: first the derivatives of 
Kolossoff variables are expressed in terms of the derivatives of the Kowalevski variables using 
Eq.(\ref{2e8}) and Eq.(\ref{2e9}). Furthermore, by substituting the conserved 
quantities Eq.(\ref{2e10}) and Eq.(\ref{2e11}), and employing the variables,  
$\{\xi_1, \xi_2, \eta_1, \eta_2\}$ expressing by the variables 
$\{x_1, x_2, \gamma_1, \gamma_2, \omega_3\}$, 
the above equations are expressed entirely in terms of the 
set $\{x_1, x_2, \gamma_1, \gamma_2, \omega_3\}$. 
By thus returning to the variables $\{x_1, x_2, \gamma_1, \gamma_2, \omega_3\}$, it has 
been confirmed that Eq.(\ref{3e11}) and Eq.(\ref{3e12}) hold true automatically.

%%%%%%%%%%%%%%%%%%%%%%%%%%%%%
\subsection{Genus Two Full Jacobi inversion Variables}
Using the variables of the genus two full Jacobi inversion problem Eq.(\ref{2e15}), 
the kinetic term of Eq.(\ref{3e11}) becomes 
as follows:
\begin{align}
&\frac{1}{k^2 (x^2+y^2) (x-k)^2}\left\{ \left((x-k)x'+y y' \right)^2 +k^2 y'^2 \right\}
\nonumber\\
&=\frac{\lambda-\mu}{\lambda+\mu}
\left( \frac{\lambda'^2}{\mu^2(\lambda^2-k^2)} 
- \frac{\mu'^2}{\lambda^2(\mu^2-k^2)} \right) .
\label{3e13}
\end{align}
Using Eqs.(\ref{2e19}), (\ref{3e11}) and (\ref{3e13}), the following equation is obtained
\begin{align}
&-\frac{c_0^2}{2}(\lambda-\mu)^2
\left( \frac{\lambda'^2}{\mu^2(\lambda^2-k^2)} 
- \frac{\mu'^2}{\lambda^2(\mu^2-k^2)} \right)
=\varphi(\lambda)-\varphi(\mu) .
\label{3e14}
\end{align}
The kinetic term in the equation of motion given by Eq.(\ref{3e12}) becomes 
as follows:
\begin{align}
&\frac{1}{k^2 (x-k)^2}
\left[ (x-k)^2 (x^2+y^2-2 k x) x'^2 +2 (x-k)y(x^2+y^2-2 k x +2 k^2) x' y' \right.
\nonumber\\
&\left. +\left\{ (y^2+k^2) (x^2+y^2-2 k x+2 k^2) -2 k^2 (x-k)^2 \right\} y'^2 \right]
\nonumber\\
&=(\lambda^2-\mu^2)^2
\left( \frac{\lambda'^2}{\mu^2(\lambda^2-k^2)} 
+ \frac{\mu'^2}{\lambda^2(\mu^2-k^2)} \right) .
\label{3e15}
\end{align}
Using Eqs.(\ref{2e22}), (\ref{3e12}) and (\ref{3e15}), the following equation is obtained
\begin{align}
&-\frac{c_0^2}{2}(\lambda-\mu)^2
\left( \frac{\lambda'^2}{\mu^2(\lambda^2-k^2)} 
+ \frac{\mu'^2}{\lambda^2(\mu^2-k^2)} \right)
=\varphi(\lambda)+\varphi(\mu) .
\label{3e16}
\end{align}
From Eq.(\ref{3e14}) and Eq.(\ref{3e16}), we obtain following equations

\begin{align}
-\frac{c_0^2}{2}(\lambda-\mu)^2
\frac{\lambda'^2}{\mu^2(\lambda^2-k^2)}=\varphi(\lambda), \quad
-\frac{c_0^2}{2}(\lambda-\mu)^2 \frac{\mu'^2}{\lambda^2(\mu^2-k^2)}
=\varphi(\mu) .
\label{3e17}
\end{align}
By choosing the sign of the square root as shown below, we obtain 
the following equation
\begin{align}
\frac{\lambda'}{\sqrt{\varphi(\lambda)(\lambda^2-k^2)}}
=-\frac{\sqrt{2} i}{c_0} \frac{\mu}{\lambda-\mu}, \quad
\frac{\mu'}{\sqrt{\varphi(\mu)(\mu^2-k^2)}}
=-\frac{\sqrt{2} i}{c_0} \frac{\lambda}{\mu-\lambda}
=\frac{\sqrt{2} i}{c_0}\frac{\lambda}{\lambda-\mu} .
\label{3e18}
\end{align}
%
%%%%%%%%%%%%%%%%%%%%%%%%%%%%%%%%%%%%%%%%%%%%%%%%%%%%%%%%%%%%%%%%%%
\section{The Genus Two Full Jacobi Inversion Problem}
\setcounter{equation}{0}

We express the fifth-degree polynomials in Eq.(\ref{3e18}) as 
$f_5(\lambda)= \varphi(\lambda)(\lambda^2 - k^2)$, which are associated 
with genus two hyperelliptic functions.
Then, Eq.(\ref{2e25}) and Eq.(\ref{3e18}) take the following form
\begin{align}
&\frac{\dot{\lambda}}{\sqrt{f_5(\lambda)}}
=\frac{\sqrt{2} i}{\lambda-\mu}, \quad
\frac{\dot{\mu}}{\sqrt{f_5(\mu)}}
=-\frac{\sqrt{2} i}{\lambda-\mu} , 
\label{4e1}\\
&\frac{\lambda'}{\sqrt{f_5(\lambda)}}
=-\frac{\sqrt{2} i}{c_0} \frac{\mu}{\lambda-\mu}, \quad
\frac{\mu'}{\sqrt{f_5(\mu)}}
=\frac{\sqrt{2} i}{c_0}\frac{\lambda}{\lambda-\mu} .
\label{4e2}
\end{align}
Then we obtain the following equation
\begin{align}
&\frac{\dot{\lambda}}{\sqrt{f_5(\lambda)}}+\frac{\dot{\mu}}{\sqrt{f_5(\mu)}}=0, \quad
\frac{\lambda'}{\sqrt{f_5(\lambda)}}+\frac{\mu'}{\sqrt{f_5(\mu)}}=\frac{\sqrt{2} i}{c_0} ,
\label{4e3}\\
&
\frac{\lambda \dot{\lambda}}{\sqrt{f_5(\lambda)}}
+\frac{\mu \dot{\mu}}{\sqrt{f_5(\mu)}}=\sqrt{2} i, \quad
\frac{\lambda \lambda'}{\sqrt{f_5(\lambda)}}
+\frac{\mu \mu'}{\sqrt{f_5(\mu)}}=0  .
\label{4e4}
\end{align}
In other words, this corresponds to the following genus two full Jacobi 
inversion problem\cite{Baker1, Baker2}
\begin{align}
&\frac{d\lambda}{\sqrt{f_5(\lambda)}}
+\frac{d\mu}{\sqrt{f_5(\mu)}}=\frac{\sqrt{2} i}{c_0} dr, \quad
\frac{\lambda d \lambda }{\sqrt{f_5(\lambda)}}
+\frac{\mu d \mu}{\sqrt{f_5(\mu)}}=\sqrt{2} i dt . 
\label{4e5}
\end{align}
Therefore, $\lambda+\mu=\wp_{22}(\sqrt{2} i r/c_0,\sqrt{2} i t)$ 
and $\lambda \mu=-\wp_{21}(\sqrt{2} i r/c_0,\sqrt{2} i t)$ give 
a solution for the two-dimensional Kowalevski top.

Finally, we compare Eq.(\ref{4e5}) with the formulation of Jacobi inversion 
problem given in the previous paper~\cite{Hayashi5}.
In the previous paper, the variables $s_1$ and $s_2$ were used instead 
of $\lambda$ and $\mu$,
and Jacobi inversion problem was expressed as follows:

\begin{align}
&\frac{d s_1}{\sqrt{f_{5,old}(s_1)}}
+\frac{d s_2}{\sqrt{f_{5,old}(s_2)}}=\frac{2 i}{c_0} dr, \quad
\frac{s_1 d s_1}{\sqrt{f_{5, old}(s_1)}}
+\frac{s_2 d s_2}{\sqrt{f_{5, old}(s_2)}}= i dt, 
\label{4e6}
\end{align}
where
\begin{align}
&f_{5,old}(s)=\varphi_{old}(s)(s^2-k^2/4), 
\nonumber\\ 
&\varphi_{old}(s)=4 s^3+2E s^2+(c_0^2-k^2) s+\frac{1}{2} E (c_0^2-k^2)-\ell^2 c_0^2.
\nonumber
\end{align}
By identifying $s_1=\lambda/2$ and $s_2=\mu/2$, and using the 
relation $\varphi(\lambda)=2 \varphi_{old}(s_1)=2 \varphi_{old}(\lambda/2)$, we 
obtain $f_{5, old}(s_1)=f_{5, old}(\lambda/2)=f_5(\lambda)/8$.
Consequently, the previous expression Eq.(\ref{4e6}) becomes equivalent to our 
expression Eq.(\ref{4e5}).

%
%%%%%%%%%%%%%%%%%%%%%%%%%%%%%%%%%%%%%%%%%%%%%%%%%%%%%%%%%%%%%%%%%%
\section{Summary and Discussions} 
\setcounter{equation}{0}

For the Kowalevski top problem, it is more natural to introduce another variable $r$, 
in addition to time variable $t$, and to formulate as the two-dimensional model.
In this paper, since the results for the genus two full Jacobi inversion 
problem for the two-dimensional Kowalevski top is already known, we utilize 
these results in reverse, introducing Kolossoff variables, to derive a more transparent 
formulation of the genus two full Jacobi inversion problem.

Since there are two implicit equations of motion using Kowalevski variables, 
we derived two explicit equations of motion using Kolossoff variables.
While, Kolossoff derived only one explicit equation using Kolosssoff variables, 
and used Hamiltonian-Jacobi theory.

The two-dimensional Kowalevski top can be formulated using a Lax pair, 
which renders this Kowalevski model an integrable system of solitons. 
Furthermore, this Kowalevski model possesses the Lie group 
structure $SO(3,2) \cong Sp(4, \mathbb{R})/\mathbb{Z}_2$~\cite{Hayashi5}. 
On the other hand, the solution to the complete genus two Jacobi problem 
is given by genus two hyperelliptic $\wp$-functions, which are integrable; 
this is because these hyperelliptic $\wp$-functions are derived from 
a single $\sigma$-function and $\log \sigma$ serves as a potential function. 
Moreover, genus two hyperelliptic $\wp$-functions 
also exhibit the Lie group structure 
$SO(3,2) \cong Sp(4, \mathbb{R})/\mathbb{Z}_2$
~\cite{Hayashi6,Hayashi7,Hayashi8} .

%%%%%%%%%%%%%%%%%%%%%%%%%%%%%%%%%%%%%%%%%

\vspace{10mm}
\noindent
{\large \bf Conflict of Interest}\\
The author declare that he has no conflicts of interest.\\

%\newpage
%%%%%%%%%%%%%%%%%%%%%%%%%%%%%%%%%%%%%%%%%%%%%%%%%%%%%%%%%%%  

\end{document}